\documentclass[12pt]{article}
 \usepackage{graphicx}
 \usepackage[cp1251]{inputenc}

\begin{document}
 \noindent {\footnotesize\it
   Astronomy Letters, 2022, Vol. 48, No. 9, pp. 542--549}
 \newcommand{\dif}{\textrm{d}}

 \noindent
 \begin{tabular}{llllllllllllllllllllllllllllllllllllllllllllll}
 & & & & & & & & & & & & & & & & & & & & & & & & & & & & & & & & & & & & & &\\\hline\hline
 \end{tabular}

  \vskip 0.5cm
  \bigskip

  \centerline {\bf Search for Close Stellar Encounters with the Solar System} 
  \centerline {\bf Based on Data from the Gaia DR3 Catalogue}

  \bigskip
  \centerline
  { V. V. Bobylev\footnote [1]{vbobylev@gaoran.ru},
   A. T. Bajkova}

 \bigskip
  \centerline{\small\it Pulkovo Astronomical Observatory, Russian Academy of Sciences,}

  \centerline{\small\it Pulkovskoe sh. 65, St. Petersburg, 196140 Russia}
 \bigskip
 \bigskip
 \bigskip
{\bf Abstract}---A search for close encounters of stars with the Solar System was performed using data from the Gaia\,DR3 catalog. We considered 31 stars with the approach parameter $d_{\rm min}<1$~pc. Among them, 15 stars are appearing as candidates for close encounters for the first time. The status of the stars GJ\,710 and HD\,7977 has been confirmed as candidates for deep penetration into the inner region of the Oort cloud. In particular, for GJ\,710 and HD\,7977, respectively, the following estimates of approach parameters are obtained:
$t_{\rm min}= 1.324\pm0.026$~Myr and $d_{\rm min}=0.052\pm0.002$~pc,
$t_{\rm min}=-2.830\pm0.025$~Myr and $d_{\rm min}=0.071\pm0.027$~pc.
Among the newly identified candidates, the most interesting is the white dwarf WD\,0810-353, for which the following approach parameters were found:
$t_{\rm min}=0.029\pm0.001$~Myr and $d_{\rm min}=0.150\pm0.003$~pc.

  \bigskip
 \section*{INTRODUCTION}
Close (less than $\sim$1~pc) encounters of field stars with the Solar System can lead to perturbation of the cometary Oort Cloud (Oort, 1950), which can provoke the emergence of the so-called comet shower from the outer boundaries of the Oort Cloud ($\sim$0.5~pc) to the inner region of the Solar System, and towards the Earth in particular. In addition to stellar flybys, the cometary Oort Cloud is subject to perturbations from giant molecular clouds, and is also affected by the Galactic tide.

The early history of searches for close encounters of stars with the Solar System using various catalogs is reflected in the works of Revina (1988),  Matthews (1994), M\"ull\"ari, Orlov (1996), Garcia-S\'anchez et al. (1999; 2001), Bobylev (2010a; 2010b) , Anderson, Francis (2012), Dybczy\'nski, Berski (2015), Bailer-Jones (2015), Feng, Bailer-Jones (2015). In particular, it was found that one of the record holders, a candidate for penetration into the Oort Cloud, is the star GJ\,710 (Garcia-Sanchez et al., 2001; Bobylev, 2010a; Berski and Dybczy\'nski, 2016).

The Gaia space experiment (Prusti et al. 2016) is dedicated to the determination of high-precision trigonometric parallaxes, proper motions, and a number of photometric characteristics of more than 1.5 billion stars. Using kinematic data on stars from various versions of this project, a large number of works have been performed on the search for close encounters of stars with the Solar System (Berski, Dybczy\'nski, 2016; Bobylev and Bajkova, 2017; Fuente Markos and Fuente Markos, 2018; Bailer-Jones et al., 2018; Darma et al., 2019; Torres et al., 2019; Wysocza\'nska et al., 2020; Bobylev and Bajkova, 2020). Thus, the analysis of data from the Gaia TGAS catalog (Tycho-Gaia Astrometric Solution, Lindegren et al. 2016) led to the discovery of several new candidates for a very close flyby, namely, for getting inside the Oort cloud to distances less than 0.5~pc. Based on the data from the Gaia\,DR2 catalog (Brown et al. 2018), it was found that approaches to the Solar System to distances less than 5~pc over the time interval $\pm$5~Ma may have about 3000 stars, less than 1~pc --- about 30~stars and 5--6~stars can have approaches to distances less than 0.25~pc.

In the Gaia\,EDR3 version (Gaia Early Data Release 3, Brown et al. 2021), the values of trigonometric parallaxes and stellar proper motions have been significantly refined compared to the previous version. In this case, the radial velocities of the stars were copied from the Gaia\,DR2 catalog. Works of Bobylev and Bajkova (2021) and Dybczy\'nski et al. (2022) are devoted to the analysis of these data in order to search for close encounters of stars with the Solar System. In particular, Dybczy\'nski et al. (2022) created an updated StePPeD electronic database (Stellar Potential Perturbers Database\footnote{https://pad2.astro.amu.edu.pl/stars/}), which contains a large amount of information about all currently known candidate stars for close encounters with the Solar System.

In the recently published version of Gaia\,DR3 (Vallenari et al. 2022), the radial velocities of stars are significantly improved --- previously measured values are refined and new ones are determined for a large number of stars. The values of trigonometric parallaxes and stellar proper motions were copied from an earlier version of the catalog, Gaia\,EDR3.

The purpose of this work is to search for close encounters of stars with the solar system using the latest data from the latest version of the Gaia~--- Gaia\,DR3 catalog.

 \section*{METHODS}\label{method}
 \subsection*{\it Linear method}
From observations, the radial velocity $V_r$ and two projections of the tangential velocity $\mu_\alpha\cos\delta$ and $\mu_\delta$ are known. The heliocentric distance to the star $d$ is calculated in terms of its trigonometric parallax $\pi,$ $d=1/\pi$. The parallax $\pi$ is expressed in milliseconds of arc (mas), and the proper motion components $\mu_\alpha\cos\delta$ and $\mu_\delta$ are expressed in milliseconds of arc per year (mas/yr). In the present work, very close stars with relative parallax errors much smaller than 10\% are used, so there is no need to take into account any corrections like the Lutz-Kelker effect (1973).

The minimum distance between the trajectories of the star and the Sun $d_{\rm min}$ at the moment of approach $t_{\rm min}$ is calculated based on the following relations (Matthews, 1994):
 \begin{equation}
 \renewcommand{\arraystretch}{1.6}
 \begin{array}{lll}
   d_{\rm min}= d/\sqrt{1+(V_r/V_t)^2},\\
   t_{\rm min}=-d\cdot V_r/(V^2_t+V^2_r),
 \label{lin}
 \end{array}
 \end{equation}
where $d$ is a heliocentric distance to the star in pc, $V_r$ is the star velocity along the line of sight and $V_t=4.74 d \sqrt{\mu^2_\alpha\cos\delta + \mu^2_\delta}$ is the star velocity perpendicular to the line of sight. Since $d$ is expressed here in pc, and the velocities $V_r$ and $V_t$ are in km/s, then $d_{\rm min}$ will be expressed in pc, and $t_{\rm min}$ in million years.

 \subsection*{\it Construction of galactic orbits of stars}
To construct the galactic orbits of stars, we use the axisymmetric gravitational potential of the Galaxy $\Phi$, which is represented as the sum of three components~--- the central spherical bulge $\Phi_b$, the disk $\Phi_d$, and the massive spherical dark matter halo $\Phi_h$:
 \begin{equation}
 \begin{array}{lll}
  \Phi=\Phi_b+\Phi_d+\Phi_h.
 \label{pot}
 \end{array}
 \end{equation}
The $\Phi_b$ bulge and $\Phi_d$ disk potentials are presented in the form proposed by Miyamoto and Nagai~(1975), and the halo component is presented according to Navarro et al.~(1997). The specific values of the parameters of the potential model used can be found in the works of Bajkova and Bobylev (2016; 2017), where it is designated as model~III.

The galactic orbits of stars are constructed in a coordinate system related to the Local Standard of Rest. Therefore, the peculiar velocity of the Sun with the values $(U,V,W)_\odot=(11.0,12.0,7.2)$~km/s from Sch\"onrich et al. (2010) is excluded from the initial velocities. The elevation of the Sun above the galactic plane $h_\odot=16$~pc is also taken into account (Bobylev and Bajkova 2016). The gravitational interaction between the star and the Sun is not taken into account.

In the work of Bobylev and Bajkova (2021), when analyzing data from the Gaia \, EDR3 catalog, the parameters of the approach of stars were calculated using the linear method, the method of integrating orbits in an axisymmetric potential, and also integrating the orbits in the potential, taking into account the spiral density wave. It was concluded that the results obtained by the second and third methods are practically the same. Therefore, in this paper, we integrate the orbits of stars without taking into account the spiral density wave.

 \begin{table}[p]                               
 \caption[]{\small\baselineskip=1.0ex\protect
 Initial data on stars
 }
 \begin{center}
 \begin{tabular}{|r|r|r|r|r|r|r|r|}\hline
 \label{tab-1}
 \def\baselinestretch{1}\normalsize\small
Gaia\,EDR3 & $\pi,$ & $\mu_\alpha\cos\delta,$ & $\mu_\delta,$ & $V_r,$ \\
                    &    mas  &  mas/yr  &  mas/yr  &  km/s  \\\hline
 4270814637616488064&$ 52.39\pm.02$&$ -0.414\pm.019$&$  -0.108\pm.017$&$ -14.42\pm0.26$\\
  510911618569239040&$ 13.21\pm.03$&$  0.144\pm.024$&$   0.010\pm.034$&$  26.76\pm0.21$\\
 5544743925212648320&$ 89.50\pm.02$&$-65.479\pm.016$&$ -29.204\pm.018$&$-373.7\pm8.2$\\
 5571232118090082816&$ 10.23\pm.01$&$  0.094\pm.010$&$   0.461\pm.011$&$  82.48\pm0.42$\\
 3372104035275483392&$ 11.97\pm.03$&$  0.557\pm.043$&$   0.438\pm.031$&$ -47.45\pm3.74$\\
  929788371508812288&$ 13.69\pm.06$&$ -2.043\pm.053$&$   3.232\pm.042$&$ 195.7\pm6.6$\\
 3118526069444386944&$  7.60\pm.05$&$  0.252\pm.049$&$   0.045\pm.049$&$  40.07\pm0.55$\\
 4116451378388951424&$ 11.86\pm.33$&$ -2.102\pm.226$&$   3.655\pm.184$&$-264.4\pm5.5$\\
 1952802469918554368&$141.89\pm.02$&$161.454\pm.017$&$-119.74\pm.019$&$ -82.91\pm0.18$\\
 3295253979286613376&$  9.33\pm.02$&$  0.261\pm.026$&$  -0.622\pm.020$&$ -61.63\pm1.89$\\
 5261593808165974784&$ 15.35\pm.01$&$ -0.091\pm.012$&$  -2.212\pm.015$&$  71.08\pm0.58$\\
 3054509410098672000&$ 24.96\pm.02$&$  0.630\pm.023$&$   5.493\pm.020$&$  70.35\pm0.35$\\
 3600338081985998080&$ 25.90\pm.09$&$ -3.175\pm.111$&$  -0.317\pm.054$&$ -37.17\pm1.95$\\
  398496965625177216&$ 13.72\pm.04$&$ -2.817\pm.046$&$  -3.043\pm.034$&$-151.4\pm7.9$\\
 4536673181955253504&$ 10.19\pm.25$&$  0.739\pm.186$&$  -0.729\pm.261$&$ -71.67\pm0.86$\\
 2926732831673735168&$ 8.85\pm.01$&$   -0.737\pm.009$&$  0.533\pm.013$&$66.46\pm0.15$\\
 3260079227925564160&$32.11\pm.03$&$   -3.615\pm.030$&$ -4.959\pm.020$&$-32.71\pm0.37$\\
 3621143693841328896&$ 7.52\pm.05$&$    0.860\pm.054$&$ -0.148\pm.046$&$85.2\pm9.0$\\
 3320184202856435840&$123.19\pm.02$&$-444.69\pm.018$&$-925.14\pm.015$&$-414.0\pm10.4$\\
 5473864079915092736&$ 9.98\pm.05$&$    8.481\pm.047$&$  9.750\pm.051$&$712.5\pm14.0$\\
 5553958176239495040&$40.27\pm.02$&$  -24.002\pm.031$&$ -0.309\pm.030$&$78.61\pm3.17$\\
 1791617849154434688&$11.38\pm.02$&$   -0.386\pm.014$&$ -1.172\pm.013$&$ 55.82\pm0.25$\\
 5346007675222666752&$44.99\pm.02$&$  -18.348\pm.017$&$-21.173\pm.016$&$ 76.29\pm0.41$\\
 6396469681261213568&$  9.80\pm.02$&$  0.490\pm.014$&$   0.188\pm.016$&$  52.22\pm0.19$\\
 6224087389269263488&$31.43\pm.15$&$   -3.235\pm.175$&$ -0.553\pm.157$&$ 18.03\pm0.45$\\
 3972130276695660288&$59.92\pm.03$&$  -20.813\pm.030$&$  6.629\pm.020$&$ 31.14\pm0.17$\\
 1926461164913660160&$316.48\pm.04$&$ 112.527\pm.036$&$-1591.6\pm.027$&$-77.29\pm0.19$\\
 5853498713190525696&$768.07\pm.05$&$-3781.7\pm.03$&$ 769.465\pm.051$&$-21.94\pm0.22$\\
 4155835025908320640&$10.50\pm.04$&$    5.892\pm.038$&$  0.062\pm.030$&$-283.0\pm10.7$\\
 2933503521200215424&$  7.43\pm.02$&$  0.248\pm.013$&$  -0.020\pm.016$&$  32.66\pm0.73$\\
 4306481867124380672&$ 9.75\pm.04$&$   -0.523\pm.024$&$  0.145\pm.022$&$-33.21\pm0.19$\\
   \hline
   \end{tabular} \end{center}
 \end{table}

 \begin{table}[p]                               
 \caption[]{\small\baselineskip=1.0ex\protect
 Additional information on stars
 }
 \begin{center}
 \begin{tabular}{|r|r|c|c|c|}\hline
 \label{tab-2}

 \def\baselinestretch{1}\normalsize\small
        Gaia\,EDR3 &   alternative designation  & StePPeD & Mass, $M_\odot$ \\\hline
4270814637616488064&                  GJ 710 & P0107 & 0.65 \\
 510911618569239040&                 HD 7977 & P0230 & 1.10 \\
5544743925212648320&             WD 0810-353 &       &      \\
5571232118090082816&    UCAC4 237-008148, AB & P0506 & 0.77 \\
3372104035275483392&                         &       &      \\
 929788371508812288&                         &       &      \\
3118526069444386944&                         & P0533 & 0.87 \\
4116451378388951424&                         &       &      \\
1952802469918554368&         LSPM J2146+3813 & P0416 & 0.20 \\
3295253979286613376&                         &       &      \\
5261593808165974784&        UCAC4 076-006432 & P0522 & 0.55 \\
3054509410098672000&                         &       &      \\
3600338081985998080&                         &       &      \\
 398496965625177216&                         &       &      \\
4536673181955253504&        UCAC4 574-064555 & P0567 & 0.60 \\
2926732831673735168&              BD-21 1529 & P0287 & 1.02 \\
3260079227925564160&        UCAC4 464-006057 & P0526 & 0.45 \\
3621143693841328896&                         &       &      \\
3320184202856435840&                         &       &      \\
5473864079915092736&                         &       &      \\
5553958176239495040&        UCAC4 213-008644 &       &      \\
1791617849154434688&         TYC 1662-1962-1 & P0189 & 0.71 \\
5346007675222666752&                         &       &      \\
6396469681261213568&          TYC 9327-264-1 & P0382 & 0.89 \\
6224087389269263488&         TYC 6760-1510-1 &       &      \\
3972130276695660288&                 GJ 3649 & P0178 & 0.55 \\
1926461164913660160&                  GJ 905 & P0413 & 0.15 \\
5853498713190525696&             Proxima Cen & P1037-C &      \\
4155835025908320640&                         &       &      \\
2933503521200215424&                HD 49995 & P0286 & 1.32 \\
4306481867124380672&               HD 179939 & P0111 & 1.69 \\
   \hline
   \end{tabular} \end{center}
 \end{table}
 \begin{table}[p]                               
 \caption[]{\small\baselineskip=1.0ex\protect
 Parameters for the approach of stars to the Solar System
 }
 \begin{center}
 \begin{tabular}{|r|rr|rr|rr|rr|}\hline
 \label{tab-3}
\def\baselinestretch{1}\normalsize\small
 Gaia\,EDR3  & $t_{\rm min},$ & $d_{\rm min},$ &
              $t_{\rm min},$ & $d_{\rm min},$ &
              $\sigma_t,$ & $\sigma_d,$\\
  &  Myr  &  pc  &   Myr  &  pc   &  Myr    &  pc       \\\hline
                     &        (1) &       &         (2) & &  &      \\\hline
 4270814637616488064&$ 1.325$&$0.051$&$ 1.324$&$ 0.052$& 0.026& 0.002\\
  510911618569239040&$-2.826$&$0.146$&$-2.830$&$ 0.071$& 0.025& 0.027\\
 5544743925212648320&$ 0.030$&$0.113$&$ 0.030$&$ 0.164$& 0.001& 0.003\\
 5571232118090082816&$-1.185$&$0.258$&$-1.185$&$ 0.203$& 0.007& 0.006\\
 3372104035275483392&$ 1.759$&$0.493$&$ 1.762$&$ 0.476$& 0.156& 0.062\\
  929788371508812288&$-0.373$&$0.494$&$-0.374$&$ 0.513$& 0.013& 0.018\\
 3118526069444386944&$-3.281$&$0.524$&$-3.288$&$ 0.529$& 0.054& 0.108\\
 4116451378388951424&$ 0.319$&$0.537$&$ 0.320$&$ 0.540$& 0.012& 0.047\\
 1952802469918554368&$ 0.084$&$0.569$&$ 0.085$&$ 0.552$& 0.001& 0.001\\
 3295253979286613376&$ 1.739$&$0.596$&$ 1.741$&$ 0.556$& 0.058& 0.022\\
 5261593808165974784&$-0.916$&$0.626$&$-0.917$&$ 0.652$& 0.008& 0.009\\
 3054509410098672000&$-0.569$&$0.598$&$-0.570$&$ 0.658$& 0.003& 0.003\\
 3600338081985998080&$ 1.038$&$0.606$&$ 1.037$&$ 0.670$& 0.060& 0.039\\
  398496965625177216&$ 0.481$&$0.689$&$ 0.482$&$ 0.730$& 0.028& 0.038\\
 4536673181955253504&$ 1.368$&$0.660$&$ 1.368$&$ 0.735$& 0.041& 0.150\\
 2926732831673735168&$-1.700$&$0.828$&$-1.701$&$ 0.786$& 0.007& 0.006\\
 3260079227925564160&$ 0.951$&$0.862$&$ 0.953$&$ 0.800$& 0.008& 0.015\\
 3621143693841328896&$-1.561$&$0.858$&$-1.558$&$ 0.809$& 0.156& 0.062\\
 3320184202856435840&$ 0.019$&$0.771$&$ 0.020$&$ 0.842$& 0.001& 0.021\\
 5473864079915092736&$-0.141$&$0.863$&$-0.142$&$ 0.857$& 0.054& 0.108\\
 5553958176239495040&$-0.315$&$0.892$&$-0.317$&$ 0.859$& 0.058& 0.022\\
 1791617849154434688&$-1.574$&$0.809$&$-1.574$&$ 0.860$& 0.026& 0.002\\
 5346007675222666752&$-0.291$&$0.859$&$-0.292$&$ 0.861$& 0.013& 0.018\\
 6396469681261213568&$-1.953$&$0.495$&$-1.949$&$ 0.872$& 0.008& 0.015\\
 6224087389269263488&$-1.763$&$0.873$&$-1.761$&$ 0.901$& 0.001& 0.001\\
 3972130276695660288&$-0.534$&$0.925$&$-0.534$&$ 0.912$& 0.060& 0.039\\
 1926461164913660160&$ 0.037$&$0.933$&$ 0.038$&$ 0.917$& 0.008& 0.009\\
 5853498713190525696&$ 0.027$&$0.958$&$ 0.028$&$ 0.939$& 0.158& 0.191\\
 4155835025908320640&$ 0.337$&$0.895$&$ 0.337$&$ 0.940$& 0.003& 0.003\\
 2933503521200215424&$-4.118$&$0.653$&$-4.113$&$ 0.980$& 0.098& 0.038\\
 4306481867124380672&$ 3.086$&$0.814$&$ 3.086$&$ 0.980$& 0.001& 0.003\\
  \hline \end{tabular} \end{center}
 {\def\baselinestretch{1}\normalsize\small {\bf Note.}
 (1)~--- linear method,
 (2)~--- axisymmetric potential.}
 \end{table}

 \subsection*{\it Estimating parameter errors}
The errors in determining $\Delta d_{\rm min}$ and $t_{\rm min}$ are estimated using the Monte Carlo method. It is assumed that the errors in the kinematic parameters of the stars are distributed according to the normal law with a root-mean-square deviation $\sigma$.

When using the linear method, measurement errors are added to the proper motion components $\mu_\alpha\cos\delta$ and $\mu_\delta$, parallax $\pi$, and radial velocity of the star $V_r$. In the method of constructing galactic orbits in the potential of the Galaxy, measurement errors are added to the rectangular coordinates $X,Y,Z$ of stars and their spatial velocities $U,V,W$.

 \section*{DATA}
First, from the catalog Gaia\,DR3, placed in the Strasbourg database under the number I/355, we selected stars using the following restrictions:
\begin{equation}
  \pi>7~\hbox{mas},~~
  \sigma_\pi/\pi<15\%,~~
  \sigma_{V_r}<15~\hbox{km/s}.
 \label{otbor}
 \end{equation}
There were about 350\,000 stars in this preliminary sample. Then we examine this sample for close approaches by two methods. Using the recommendations made in Bobylev and Bajkova (2021), we first select candidates based on a simpler linear method, then we apply a more complex second method for estimating the approach parameters using the gravitational potential of the Galaxy to the selected stars.

 \section*{RESULTS}
Tables~\ref{tab-1}--\ref{tab-3} show the results of the search for candidate stars for close encounters with the Solar System. Table~\ref{tab-1} gives the initial data on stars extracted from the Gaia\,DR3 catalog.

Table~\ref{tab-2} gives the numbers of selected stars from the Gaia\,EDR3 catalog and their alternative designations, as well as numbers from the StePPeD database (Dybczy\'nski et al., 2022) and mass estimates taken from the StePPeD database. Note that for many stars in Table~\ref{tab-2} there are no alternative designations. Moreover, most of these stars are missing from the SIMBAD~\footnote{https://simbad.u-strasbg.fr/simbad/} search database. That is, the complete set of kinematic parameters ($\pi, \mu_\alpha\cos\delta, \mu_\delta$, and $V_r$) necessary for analysis for such stars, appeared for the first time.

Table~\ref{tab-3} gives the parameters of the approach of stars to the Solar System $t_{\rm min}$ and $d_{\rm min}$ calculated by two methods. In tables~\ref{tab-1}--\ref{tab-3} the stars are sorted by the parameter $d_{\rm min}$, found when constructing orbits using the axisymmetric potential (method (2) in Table~\ ref{tab-3}). The Tables give 31 stars with $d_{\rm min}<1$~pc.

 \section*{DISCUSSION}
As can be seen from Table~\ref{tab-2}, the masses of candidates are small. The magnitude of the perturbation that it can exert on the cometary cloud strongly depends on the mass of a passing star. The most massive star of our list occupies the last line in the tables. The outer boundary of the cometary Oort cloud is not exactly known. It is assumed that it is about 0.48~pc. In our opinion, 5-6 stars, in which the value of the parameter $d_{\rm min}$ is inside the outer boundary of the Oort cloud, may be of the greatest interest. We will discuss them in more detail.

{\bf GJ\,710 and HD\,7977.} The first two stars in Tables~\ref{tab-1}--\ref{tab-3} and earlier, when using data from the Gaia\,EDR3 catalog, occupied the first two lines of candidates for the closest encounters (Bobylev and Bajkova, 2021). We note that the random errors in the estimates of the parameters $t_{\rm min}$ and $d_{\rm min},$ have approximately halved, which is due to improved radial velocities for these stars in the Gaia\,DR3 catalog.

Fuente Marcos and Fuente Marcos (2022) estimated the parameters of the approach of the star GJ\,710 using data from the Gaia\,DR3 catalog. These authors have shown that the distribution of the distances of the minimum approach to the Solar System has a median value of 0.052 pc, and with a probability of 90\% is in the range of 0.048--0.056 pc; the corresponding perihelion passage time lies in the range of 1.26 and 1.33 million years with a confidence of 90\%, with the most likely value of 1.29 million years. The values of these parameters found by us are in good agreement with the estimates of Fuente Marcus and Fuente Marcus.

{\bf WD\,0810-353} ranks third in the tables~\ref{tab-1}--\ref{tab-3}. This star appears for the first time in the list of candidates for the closest approaches to the solar system. According to the SIMBAD database, this is a white dwarf with large proper motions, therefore it is classified as HPMS (High Proper Motion Star). According to Bagnulo and Landstreet (2020), this white dwarf has a mass of $0.63 M_\odot$ and an age of 2.7 Gyr.

This star also has a huge radial velocity $V_r=-373.7\pm8.2$~km/s. For white dwarfs, it is important to take into account the gravitational redshift of spectral lines when determining the radial velocities. Such an effect is a consequence of a strong gravitational field on the surface of white dwarfs, whose influence is equivalent to velocities of several tens of km/s (see, for example, Greenstein and Trimble (1967)). We performed new calculations with the radial velocity value $V_r=-423.7\pm8.2$~km/s, adding $-50$~km/s to the measured value, and found the following approach parameters: $t_{\rm min}= 0.029\pm0.001$~million years and $d_{\rm min}=0.150\pm0.003$~pc. In this case, taking into account the correction for the gravitational redshift does not have a critical effect on the estimation of the rendezvous parameters.

In general, we can conclude that in about 30 thousand years the star will rapidly sweep past the solar system. Its influence on the objects of the Oort cloud will be very short.

{\bf UCAC4 237-008148.} This low-mass binary star is number four in the tables~\ref{tab-1}--\ref{tab-3}. For example, in the work of Dybczy\'nski et al. (2022), when analyzing data from the Gaia\,EDR3 catalog, the following approach parameters were found for it:
$t_{\rm min}=-1.084\pm0.004$~Myr and $d_{\rm min}=0.199\pm0.008$~pc. We can see that there is good agreement with the data in Table~\ref{tab-3}, especially good agreement is in the estimation of parameter errors.

{\bf Gaia\,EDR3 3372104035275483392 and Gaia\,EDR3 929788371508812288.} These two stars rank fifth and sixth in our tables.  They appear for the first time in the list of candidates for the closest approaches to the solar system. Unfortunately, they are not in the SIMBAD database, so little is known about these stars.

In general, we can conclude that due to the use of selection criteria~(\ref{otbor}), the approach parameters of all stars are determined reliably, since the errors in determining these parameters are small (see the last two columns of Table~\ref{tab-3}) .

 \section*{CONCLUSION}
A search for close encounters of stars with the Solar System was carried out using data from the latest Gaia\,DR3 catalog. The paper considers 31 stars with the approach parameters $d_{\rm min}<1$~pc. Among them, 15 stars for the first time act as candidates for close encounters with the solar system. These 15 stars were identified mainly due to the appearance of measurements of their radial velocities in the Gaia\,DR3 catalog.

For the stars GJ\,710 and HD\,7977, their status as candidates for very close encounters, candidates for deep penetration into the inner region of the Oort cloud, has been confirmed.

Among the newly identified candidates, the most interesting is the white dwarf WD\,0810-353. For this star, the following approach parameters were found, taking into account the model correction for the effect of gravitational redshift in spectral lines when measuring its radial velocity:
$t_{\rm min}=0.029\pm0.001$~Myr and $d_{\rm min}=0.150\pm0.003$~pc.

 \bigskip\medskip{REFERENCES}\medskip{\small
 \begin{enumerate}

 \item
E. Anderson, et al., Astron. Lett. {\bf 38}, 331 (2012).

 \item
S. Bagnulo and J.D. Landstreet, Astron. Astrophys. {\bf 643}, 134 (2020).

 \item
C.A.L. Bailer-Jones, Astron. Astrophys. {\bf 575}, 35 (2015).

 \item
C.A.L. Bailer-Jones, J. Rybizki, R. Andrae, and
 M. Fouesneau, Astron. Astrophys. {\bf 616}, 37 (2018). 

 \item
A.T. Bajkova and V.V. Bobylev, Astron. Lett. {\bf 42}, 567 (2016). 

 \item
A.T. Bajkova and V.V. Bobylev, Open Astron. {\bf 26}, 72 (2017).

 \item
F. Berski and P.A. Dybczy\'nski, Astron. Astrophys. {\bf 595}, L10 (2016).

 \item
V.V. Bobylev, Astron. Lett. {\bf 36}, 220 (2010a). 

 \item
V.V. Bobylev, Astron. Lett. {\bf 36}, 816 (2010b). 

 \item
V.V. Bobylev and A.T. Bajkova, Astron. Lett. {\bf 42}, 1 (2016). 

 \item
V.V. Bobylev and A.T. Bajkova, Astron. Lett. {\bf 43}, 559 (2017). 

 \item
V.V. Bobylev and A.T. Bajkova, Astron. Lett. {\bf 46}, 245 (2020). 

 \item
V.V. Bobylev and A.T. Bajkova, Astron. Lett. {\bf 47}, 180 (2021). 

 \item
Gaia Collaboration, A.G.A. Brown, A. Vallenari, T. Prusti, et al.,
Astron. Astrophys. {\bf 616}, 1 (2018). 

 \item
Gaia Collaboration, A.G.A. Brown, A. Vallenari, T. Prusti, et al.,
Astron. Astrophys. {\bf 649}, 1 (2021). 

 \item
R. Darma, W. Hidayat, and M.I. Arifyanto, J. Phys.: Conf. Ser. {\bf 1245}, 012028 (2019).

 \item
P.A. Dybczy\'nski and F. Berski, MNRAS {\bf 449}, 2459 (2015).

 \item
P.A. Dybczy\'nski, F. Berski, J. Tokarek, et al., arXiv: 2206.1104 (2022).

\item
F. Feng and C.A.L. Bailer-Jones, MNRAS {\bf 454}, 3267 (2015).

\item
R. de la Fuente Marcos and C. de la Fuente Marcos,
 Res. Not. Am. Astron. Soc. {\bf 2}, 30 (2018).

\item
R. de la Fuente Marcos and C. de la Fuente Marcos,
 Res. Not. Am. Astron. Soc. {\bf 6}, 136 (2022).

 \item
J. Garcia-S\'anchez, R.A. Preston, D.L. Jones, et al.,
Astron. J. {\bf 117}, 1042 (1999).

 \item
J. Garcia-S\'anchez, P.R. Weissman, R.A. Preston, et al.,
Astron. Astrophys. {\bf 379}, 634 (2001).

 \item
J.L. Greenstein and V.L. Trimble, Astrophys. J. {\bf 149}, 283 (1967).

 \item
Gaia Collaboration, L. Lindegren, U. Lammers, U. Bastian, et al.,
Astron. Astrophys. {\bf 595}, A4 (2016). 

 \item
T.E. Lutz and D.H. Kelker, PASP {\bf 85}, 573 (1973).

 \item
M. Miyamoto and R. Nagai, PASP {\bf 27}, 533 (1975).

 \item
R.A.J. Matthews, Royal Astron. Soc. Quart. Jorn. {\bf 35}, 1 (1994).

 \item
A.A. M\"ull\"ari and V.V. Orlov, {\it Earth, Moon, and Planets} (Kluwer, Netherlands, {\bf 72}, p. 19, 1996).

 \item
J.F. Navarro, C.S. Frenk and S.D.M. White, Astrophys. J. {\bf 490}, 493 (1997).

 \item
J.H. Oort, Bull. Astron. Inst. Netherland {\bf 11}, No 408, 91 (1950).

\item
Gaia Collaboration,  T. Prusti, J.H.J. de Bruijne, A.G.A. Brown, et al.,
Astron. Astrophys. {\bf 595}, 1 (2016).

\item
I.A. Revina, {\it Analysis of motion of celestial bodies  and estimation of accuracy of their observations} (Latvian University, Riga, p.~121, 1988).

\item
S. Torres, M.X. Cai, A.G.A. Brown, and S. Portegies Zwart,
Astron. Astrophys. {\bf 629}, 139 (2019).

 \item
R. Sch\"onrich, J. Binney and W. Dehnen, MNRAS {\bf 403}, 1829 (2010).

 \item
Gaia Collaboration, A. Vallenari, A.G.A. Brown, T. Prusti, et al., in press (2022).

 \item
R. Wysocza\'nska, P.A. Dybczy\'nski, and M. Poli\'nska,
Astron. Astrophys. {\bf 640}, 129 (2020). 

 \end{enumerate}
\end{document}